\documentstyle[11pt,moriond,epsfig]{article}

\def\Journal#1#2#3#4{{#1} {\bf #2}, #3 (#4)}

\def\NIMA{{\em Nucl. Instrum. Methods} A}

\def\PLB{{\em Phys. Lett.}  B}
\def\PRL{\em Phys. Rev. Lett.}
\def\PRD{{\em Phys. Rev.} D}

\def\be{\begin{equation}}
\def\ee{\end{equation}}
\def\bea{\begin{eqnarray}}
\def\eea{\end{eqnarray}}

\begin{document}
\vspace*{4cm}
\title{FIRST MEASUREMENT OF $\Gamma(D^{\ast +})$}

\author{ M. DUBROVIN }

\address{Wilson Laboratory, Cornell University / Wayne State University \\
Ithaca, NY 14953, USA \\ (on behalf of CLEO Collaboration) }

\maketitle\abstracts{
We have made the first measurement of the $D^{\ast +}$ width using
9/fb of $e^+e^-$ data collected near the $\Upsilon(4\rm{S})$
resonance by the CLEO II.V detector.
Our method uses advanced tracking techniques
and a reconstruction method that takes advantage of the small
vertical size of the CESR beam spot to measure the energy release
distribution from the $D^{\ast +} \to D^0 \pi^+$ decay.
Our preliminary result is $\Gamma(D^{\ast +}) = 96 \pm 4\ ({\rm Statistical})
\pm 22\ ({\rm Systematic})$ keV. }

%\section{Introduction}

A measurement of $\Gamma(D^{\ast +})$ opens an important window
on the non-perturbative strong physics involving heavy quarks.
The basic framework of the theory is well understood, however, there is
still much speculation -
predictions for the width range from $15\,\rm{keV}$ to 
$150\,\rm{keV}$ \cite{pred}.
We know the $D^{\ast +}$ width is dominated by strong decays,
since the measured magnetic-dipole transition rate is small,
$Br(D^{*+}\to D^+\gamma) = 1.68\pm0.45\%$, \cite{mats} 
and can be neglected to the first order.
The level splitting in the $B$ sector
is not large enough to allow real strong transitions.
Therefore, a measurement of the $D^{\ast +}$ width
gives unique information about the
strong coupling constant in heavy-light systems \cite{Wise}, 
$g_{D^*D\pi}$ or $g$.

Prior to this measurement, the $D^{\ast +}$ width was limited
to be less than $131\,\rm{keV}$ at the $90\%$ confidence level 
by the ACCMOR collaboration \cite{ACCMOR}.
The limit was based on 110 signal events reconstructed in two
$D^0$ decay channels with a background of 15\% of the signal.
This contribution describes
a measurement of the $D^{\ast +}$ width with the CLEO II.V detector
where the signal, in excess of 11,500 events,  is
reconstructed through a single, well-measured sequence,
$D^{\ast +} \rightarrow \pi^+_{\rm slow} D^0$,
$D^0 \rightarrow K^-\pi^+$.  Consideration of
charge conjugated modes are implied throughout this paper.
The level of background under the signal is less than 3\%
in our loosest selection.

	The CLEO detector has been described in detail elsewhere.  All of
the data used in this analysis are taken with the detector in its
II.V configuration \cite{CLEO}.  
	The data were taken in symmetric $e^+e^-$ collisions at a center of
mass energy around 10 GeV with an integrated luminosity of 9.0/fb provided by
the Cornell Electron-positron Storage Ring (CESR).  The nominal sample
follows the selection of
$D^{\ast +} \to \pi_{\rm slow}^+ D^0 \to \pi_{\rm slow}^+ K^-\pi^+$ candidates
used in our $D^0-\bar{D^0}$ mixing analysis.\cite{Dmix}

\begin{figure}[!htb]
  \begin{minipage}[t]{75mm}
  \epsfxsize=90mm
  \centerline{\epsfbox{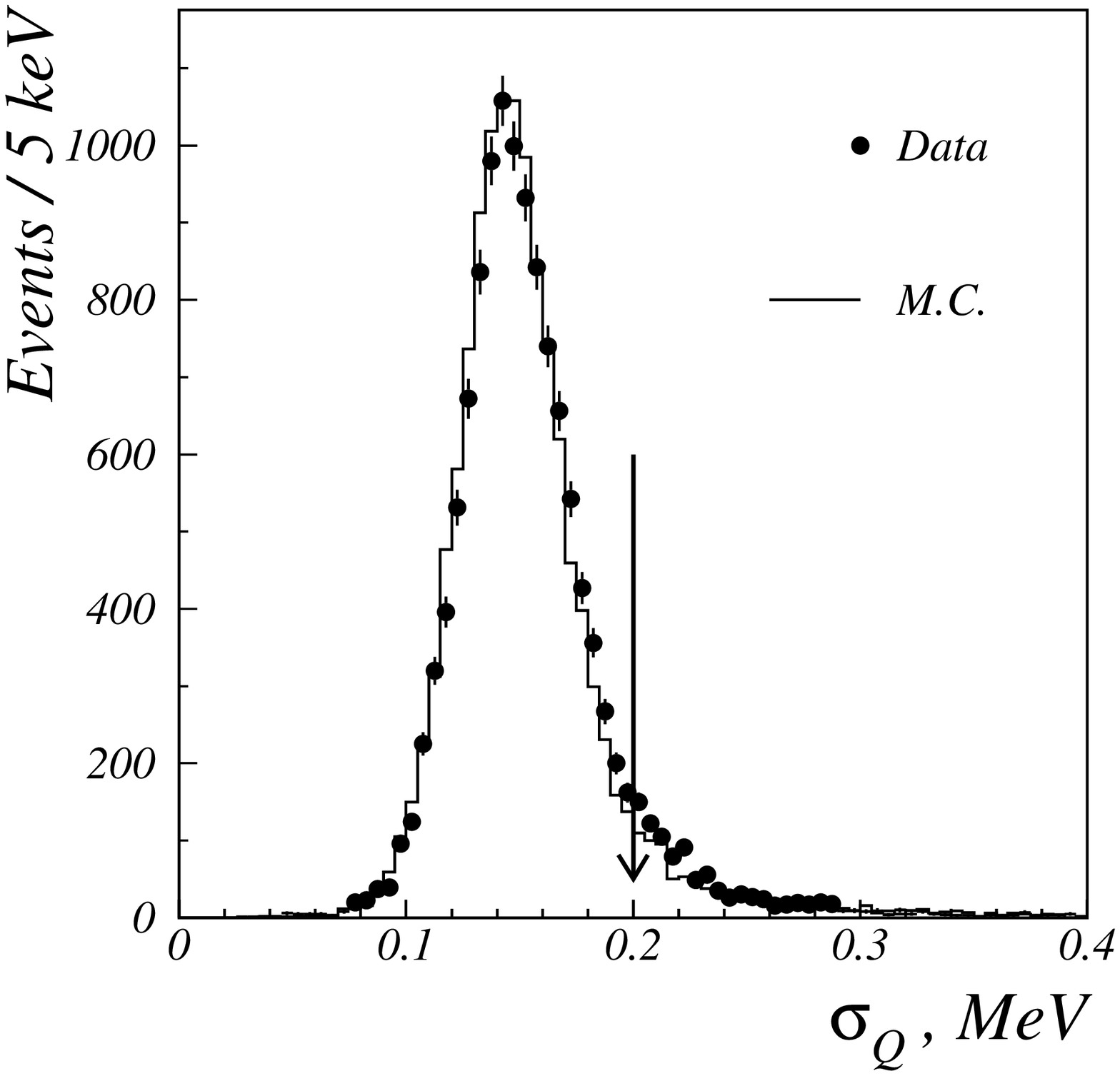}}
     \caption{\label{fig:errorcompare} Distribution of $\sigma_Q$,
              the uncertainty on $Q$ as determined from propagating
              track fitting errors.  The arrow indicates a selection cut.}
  \end{minipage}
\hfill
  \begin{minipage}[t]{75mm}
  \epsfxsize=90mm
  \centerline{\epsfbox{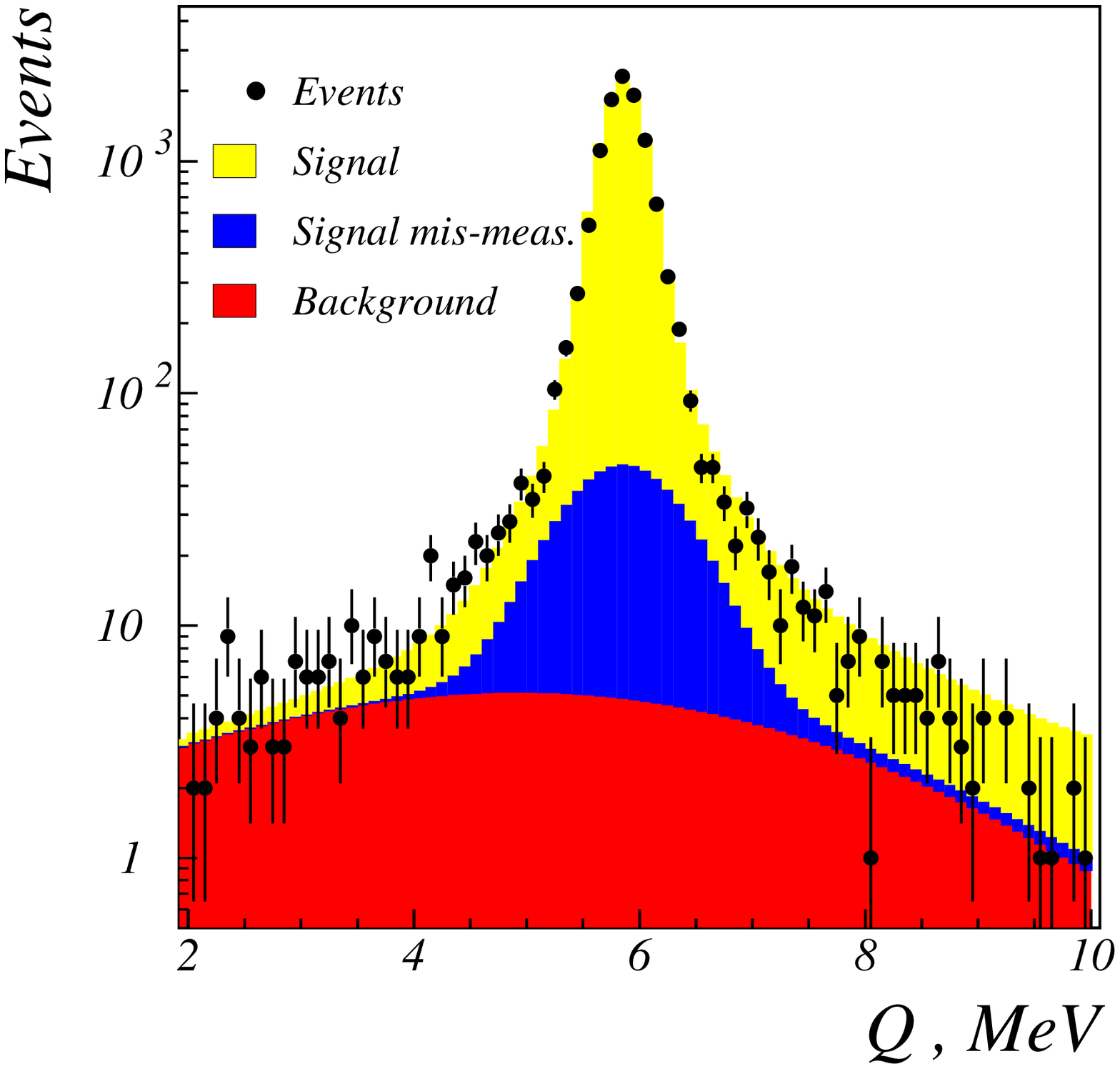}}
  \caption{\label{fig:nomfit} Fit to {\em nominal} data sample.  
  The different contributions to the fit are shown by different colors.}
  \end{minipage}
  \begin{minipage}[t]{75mm}
  \epsfxsize=90mm
  \centerline{\epsfbox{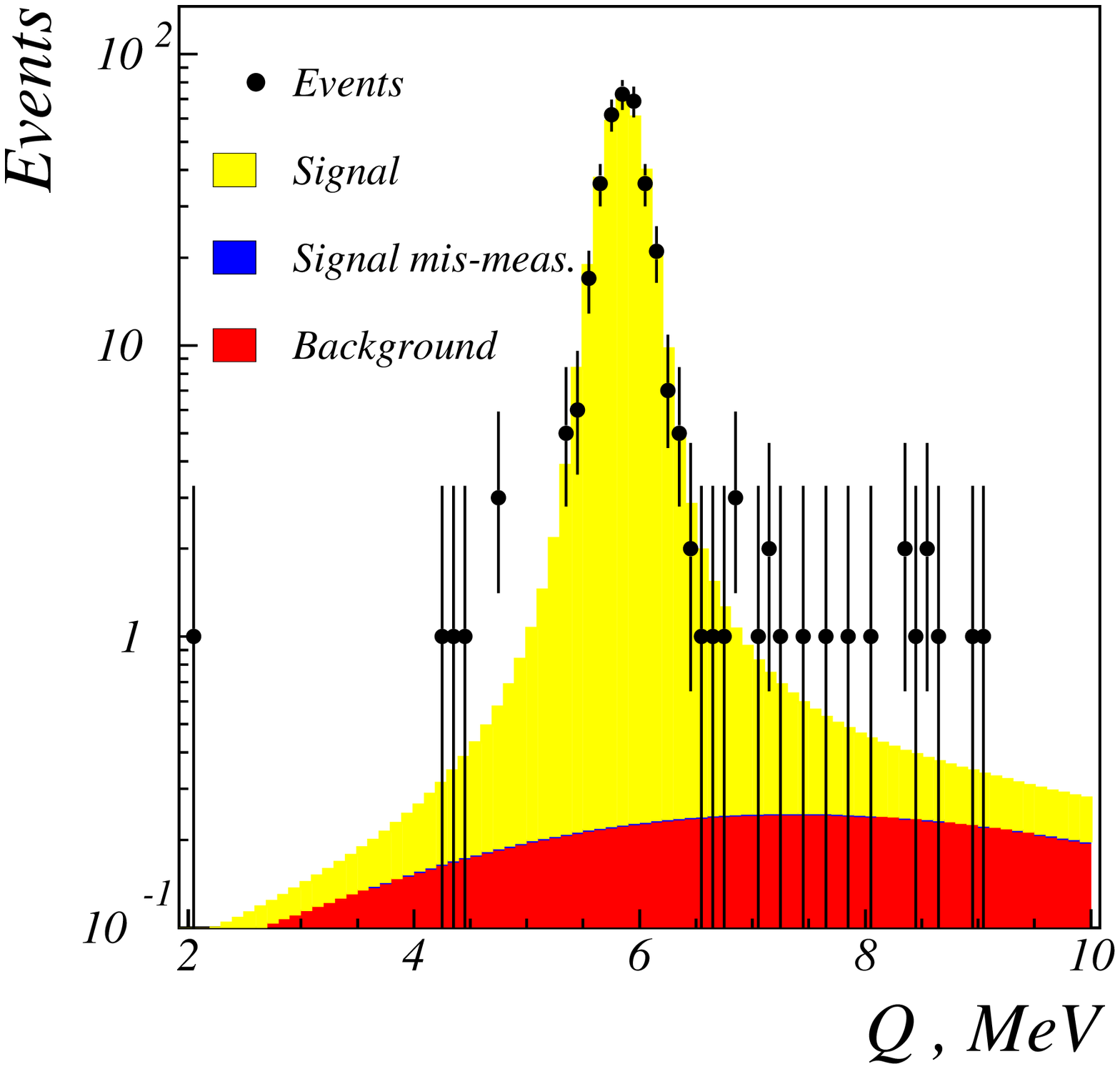}}
  \caption{\label{fig:platinumfit} Fit to {\em tracking selected} data sample.}
  \end{minipage}
\hfill
  \begin{minipage}[t]{75mm}
  \epsfxsize=90mm
  \centerline{\epsfbox{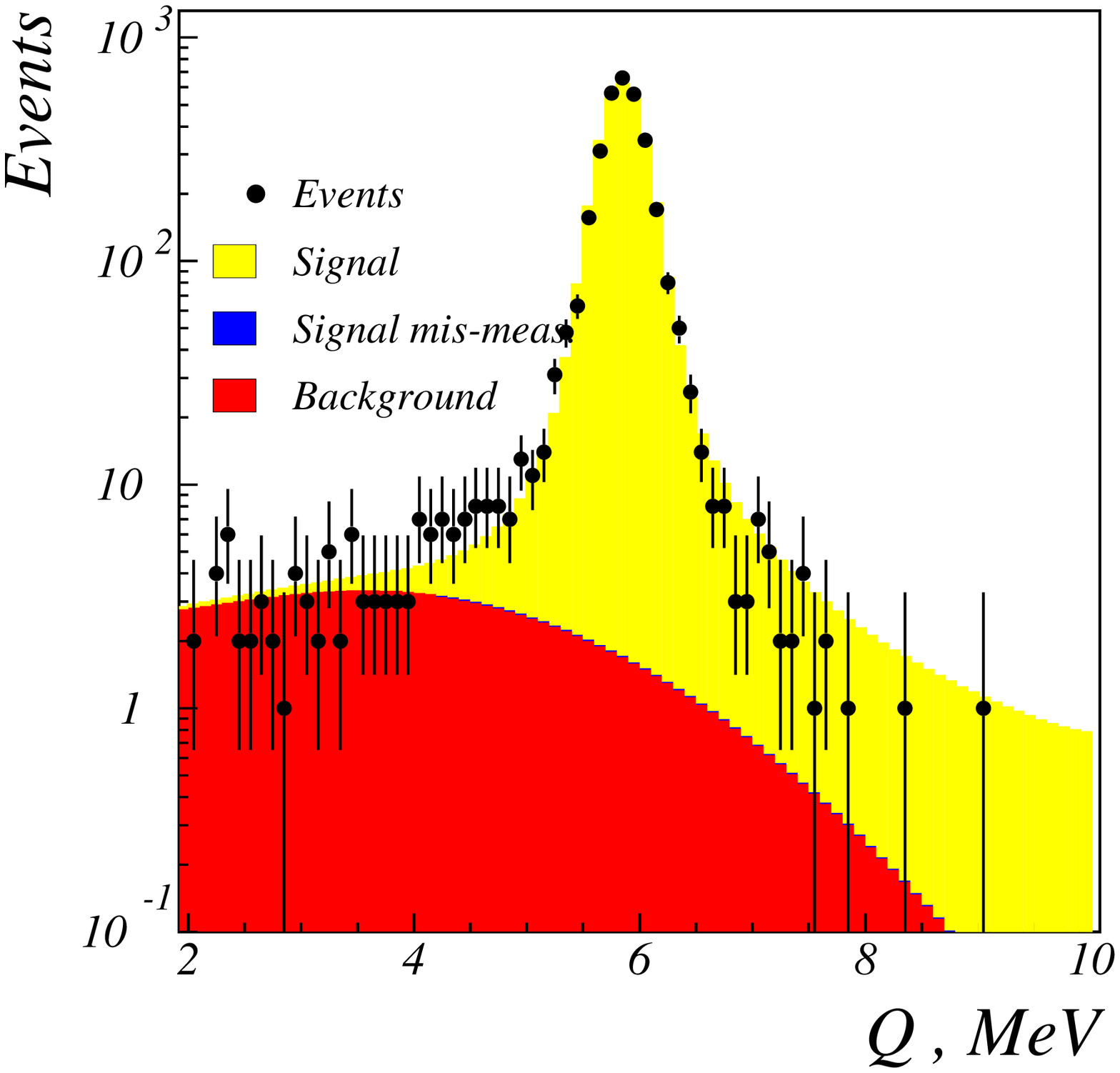}}
  \caption{\label{fig:goldfit} Fit to {\em kinematic selected} data sample.}
  \end{minipage}
\end{figure}

	Our reconstruction method takes advantage of the small CESR beam
spot and the kinematics and topology of the
$D^{\ast +} \to \pi^+_{\rm slow} D^0 \to \pi^+_{\rm slow} K^- \pi^+$
decay chain. The $K^-$ and $\pi^+$ are required to form a common vertex.
The resultant $D^0$ candidate momentum vector is then projected back to the
CESR luminous region to determine the $D^0$ production point.  
The CESR luminous region has a Gaussian width $\sim 10\ \mu$m vertically and
$\sim 300\ \mu$m horizontally.  
This procedure determines an accurate $D^0$ production point.
Then the $\pi_{\rm slow}^+$ track
is re-fit constraining its trajectory to intersect the $D^0$
production point.  This improves the resolution on the energy release,
$Q$, by more
than 30\% over simply forming the appropriate invariant masses of the tracks.
The improvement to resolution is essential to our measurement of the width
of the $D^{\ast +}$.  Our resolution is shown in Figure~\ref{fig:errorcompare}
and is typically 150 keV.
The good agreement reflects that the kinematics
and sources of errors on the tracks, such as number of hits
and the effects of multiple scattering in detector material, 
are well modeled.

The challenge of measuring of the $D^{\ast +}$ width 
is understanding the tracking system response function since 
the experimental resolution exceeds the width we are trying to measure. 
We depend
on exhaustive comparisons between a GEANT \cite{GEANT} based detector
simulation and our data.  We addressed the problem by selecting samples of
candidate $D^{\ast +}$ decays using three strategies.
	First we produced the largest sample from data and simulation by 
imposing only
basic tracking consistency requirements.  We call this the
{\em nominal} sample.
	Second we refine the nominal sample selecting candidates
with the best measured tracks by making very tight cuts
on tracking parameters.  There is special emphasis on choosing those tracks
that are well measured in our silicon vertex detector.  This reduces
our nominal sample by a factor of thirty and, according to our
simulation, has negligible contribution from tracking mishaps.
We call this the {\em tracking selected} sample.
	A third alternative is to select our data on specific kinematic
properties of the $D^{\ast +}$ decay that minimize the dependence
of the width of the $D^{\ast +}$ on detector mismeasurements.
The nominal sample size is reduced by a factor
of three and a half and, again according to our simulation,
the effect of tracking problems is reduced to negligible levels.
We call this the {\em kinematic selected} sample.
	Table~\ref{tab:data} 
summarizes the statistics in our three samples.

	In all three samples the width is extracted with an unbinned maximum
likelihood fit to the energy release, 
$Q = m(K^-\pi^+\pi_{\rm slow}^+) - m(K^-\pi^+) - m_{\pi^+}$, 
distribution and compared with
the simulation's generated value to determine a bias which is
then applied to the data.
The three different samples yield consistent values for
the width of the $D^{\ast +}$ giving us confidence that our simulation
accurately models our data.

We assume that the intrinsic width of the $D^0$ is negligible,
$\Gamma(D^0) \ll \Gamma(D^{\ast +})$, 
implying that the observed width of $Q$ distribution
is dominated by the shape given by the $D^{\ast +}$ intrinsic width and
the tracking system response function.  Thus we
consider the pairs of $Q_i$ and $\sigma_{Qi}$ for
$D^{\ast +} \to \pi_{\rm slow}^+ D^0 \to \pi_{\rm slow}^+ K^-\pi^+$
where $\sigma_{Qi}$ is given for
each $i$-th candidate by propagating the tracking errors in the kinematic
fit of the charged tracks \cite{kalman}.  
We perform an unbinned maximum likelihood fit
to the $Q$ distribution, minimizing the likelihood function
\begin{equation}
L = 2 (N_s + N_b) - 2 \sum _{i=1}^N \log 
[N_s \cdot S(Q_i, \sigma_{Qi}; \Gamma_0, Q_0, f_{mis}, \sigma_{mis} ) + 
N_b \cdot B(Q_i; b_{1,2,3})],
\end{equation}
where $S$ and $B$ are respectively the signal and the background shapes,
$N_s$ and $N_b$ number of signal and background events.

The shape of the underlying signal is assumed to be
given by a P-wave Breit-Wigner,
with central value of $Q$, $Q_0$.
We considered a relativistic and a non-relativistic
Breit-Wigners as a model of the
underlying signal shape, and found negligible difference in the fit
parameters.
The width of the Breit-Wigner depends on $Q$ and is given by
\begin{equation}
\Gamma(Q) = \Gamma_0  \left(\frac{P}{P_0}\right)^3 
\left(\frac{m_0}{m}\right)^2 ,
\label{eq:BW}
\end{equation}
where $\Gamma_0$ is equivalent to $\Gamma(D^{\ast +})$,
$m$ and $P$ are the measured candidate $D^{\ast +}$ mass
and $\pi_{\rm slow}^+$ or $D^0$ momentum in
the $D^{\ast +}$ rest frame and $P_0$ and $m_0$ are
the values computed using $Q_0$.
The effect of the mass term is negligible at our energy. 
The partial width and the total width differ negligibly in  
their dependence on $Q$ for $Q>1~MeV$. We use Eqn~\ref{eq:BW}
suitably normalized to describe of $Q$ effects on width.

For each candidate the signal shape, $S$, is a convolution
of the Breit-Wigner function with
a resolution Gaussian with width, $\sigma_{Qi}$, determined by the tracking
errors, as a model of  our finite resolution.  Figure~\ref{fig:errorcompare}
shows the distribution of $\sigma_Q$ for the data and the simulation.
We allow a small fraction of the signal, $f_{mis}$, to be parameterized
by a single Gaussian with effective resolution, $\sigma_{mis}$,
different from measured, $\sigma_{Qi}$.
This shape is included in the fit to model the tracking
mishaps, mis-assigned hits and hard multiple scatters, 
which our simulation predicts to be at the 5\% level in the
nominal sample and negligible in both the tracking
and kinematic selected samples.
For the purpose of systematic study
the $\sigma_Q$ has a scale factor, $k$, which is fixed to one in 
our nominal fits. 

The fit also includes a background contribution with fixed shape,
$B$, presented by polynomial function
with three or more parameters, $b_{1,2,3}$. 
This shape is taken from fits to the
background prediction of our simulation.
The level of the background is allowed to float in our standard fit.

	The fitter has been extensively tested both numerically
and with input from our full simulation.  We find that the fitter
performs reliably giving normal distributions for the floating
parameters and their uncertainties.  It also reproduces the input
$\Gamma(D^{\ast +})$ from 0 to 130 keV with offset consistent 
with zero, as shown in Table~\ref{tab:data}.

We note that if all the parameters are allowed to vary simultaneously
there is strong correlation among the intrinsic width $\Gamma_0$,
the fraction of mismeasured events $f_{mis}$, and the $\sigma_Q$ scale
factor $k$, as one would expect.
Thus our nominal fit holds $k$ fixed to one, but in our systematic
studies we either fix one of the three or provide a constraint with
a contribution to the likelihood if the parameter varies from its
nominal value.

	Figures~\ref{fig:nomfit}, \ref{fig:platinumfit}, and \ref{fig:goldfit}
respectively display the fit to the nominal, tracking, and kinematic selected
data samples.  The results of the fits are summarized in Table~\ref{tab:fit}.
Correlations among the floating parameters of the fit are negligible.

\begin{table}
\caption{Summary of our data sample, simulation biases, and fit results.}
\begin{center}
\begin{tabular}{|c|c|c|c|} \hline
                 & \multicolumn{3}{|c|}{Sample} \\ \cline{2-4}
 Parameter       & Nominal          & Tracking          & Kinematic \\ \hline
 Candidates           & 11496            & 368               & 3284 \\
 Background Fraction (\%)
                     & $2.51 \pm 0.27$  & $4.1 \pm 1.9$     & $4.05 \pm 0.49$ \\
 $\Gamma_{\rm fit} - \Gamma_{\rm generated}$ (keV)
                     & $2.7 \pm 2.1$    & $1.7 \pm 6.4$     & $4.3 \pm 3.1$ \\
 Fit $\Gamma_0$ (keV) &  $98.9 \pm 4.0$  & $106.0 \pm 19.6$ & $108.1 \pm  5.9$ \\       
 $D^{\ast +}$ Width (keV) & $96.2 \pm 4.0$
                         & $104 \pm 20$
                         & $103.8 \pm 5.9$ \\ \hline
\end{tabular}
\end{center}
\label{tab:data}
\end{table}

\begin{table}
\caption{Results of the fits described in the text.
 The uncertainties are statistical.}
\begin{center}
\begin{tabular}{|c|c|c|c|} \hline
                      & \multicolumn{3}{|c|}{Sample} \\ \cline{2-4}
Parameter            & Nominal          & Tracking          & 
Kinematic \\ \hline
$\Gamma_0$ (keV)     & $98.9 \pm 4.0$   & $106.0 \pm 19.6$  & $108.1 
\pm  5.9$ \\
$Q_0$ (keV)          & $5853 \pm 2$     & $5854 \pm 10$     & $ 5850 \pm 4$ \\
$N_s$                & $11207 \pm 109$  & $353 \pm 20$      & $3151 \pm 57$ \\
$f_{mis}$ (\%)       & $5.3 \pm 0.5$    & NA                & NA \\
$\sigma_{mis}$ (keV) & $508 \pm 39 $    & NA                & NA \\
$N_b$                & $289 \pm 31 $    & $15 \pm 7$        & $133 \pm 16$ \\ \hline
\end{tabular}
\end{center}
\label{tab:fit}
\end{table}

\begin{figure}
\begin{tabular}{ccc}
\epsfxsize=55mm
\epsfbox{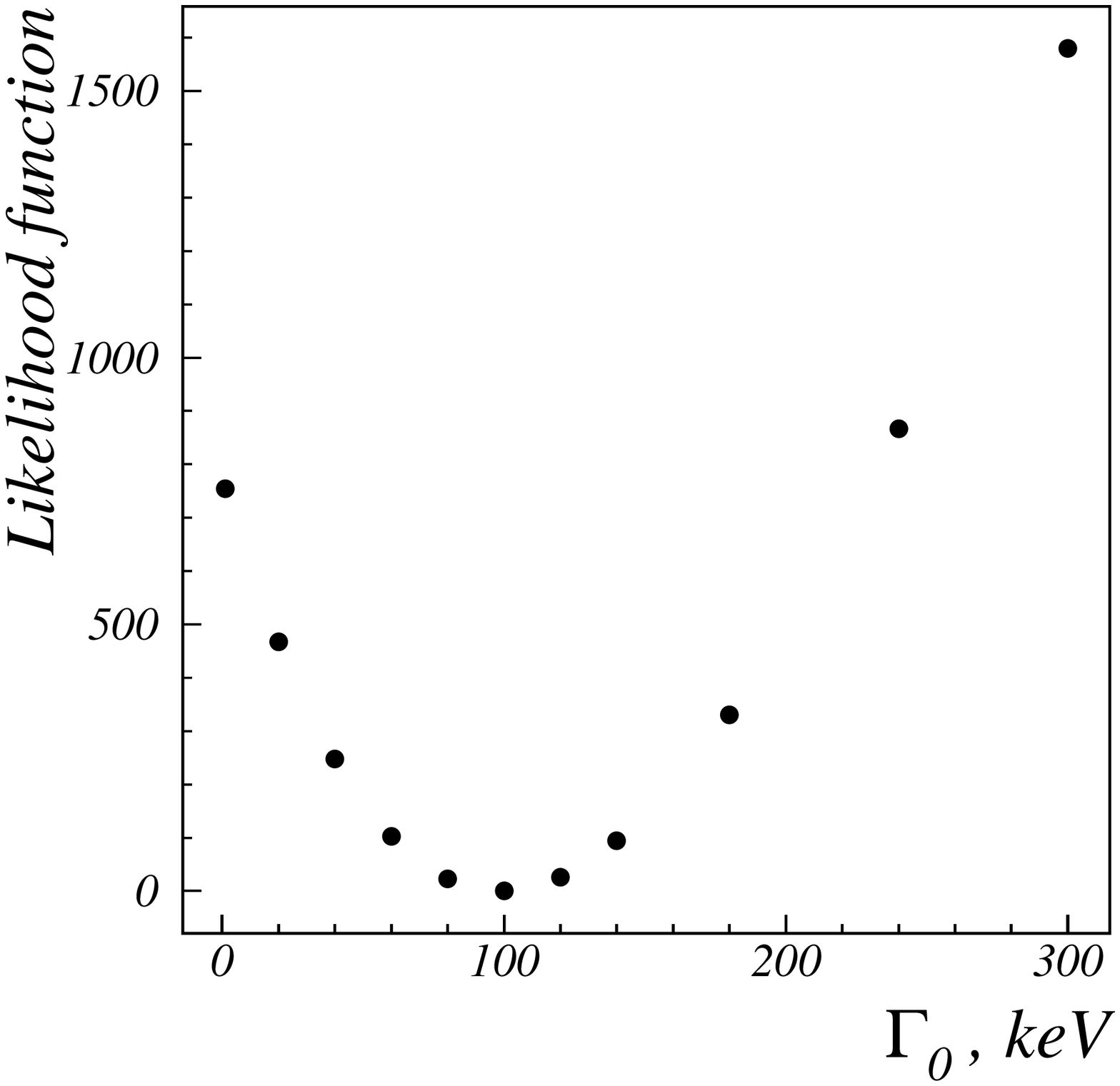} &
\epsfxsize=55mm
\epsfbox{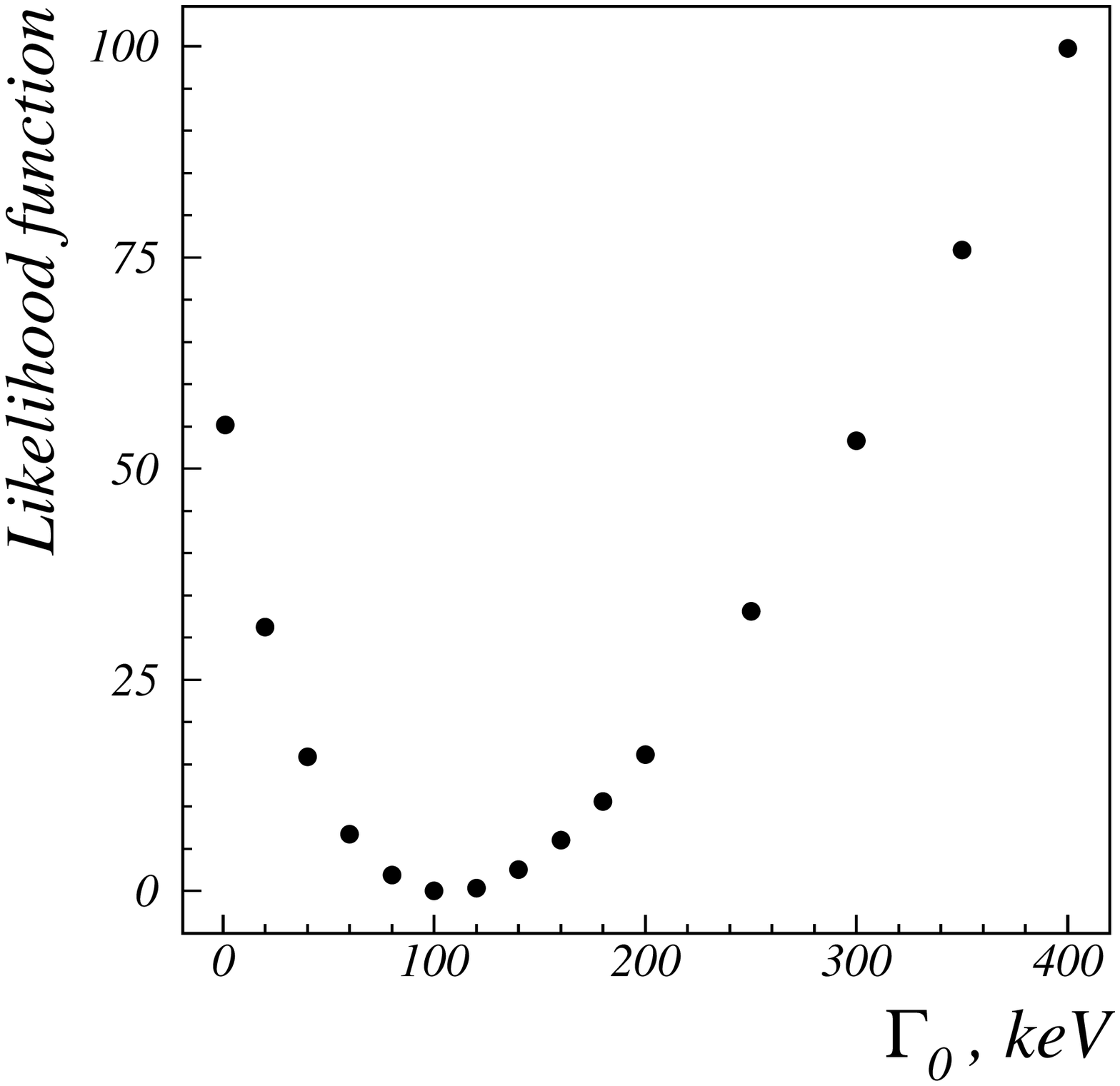} &
\epsfxsize=55mm
\epsfbox{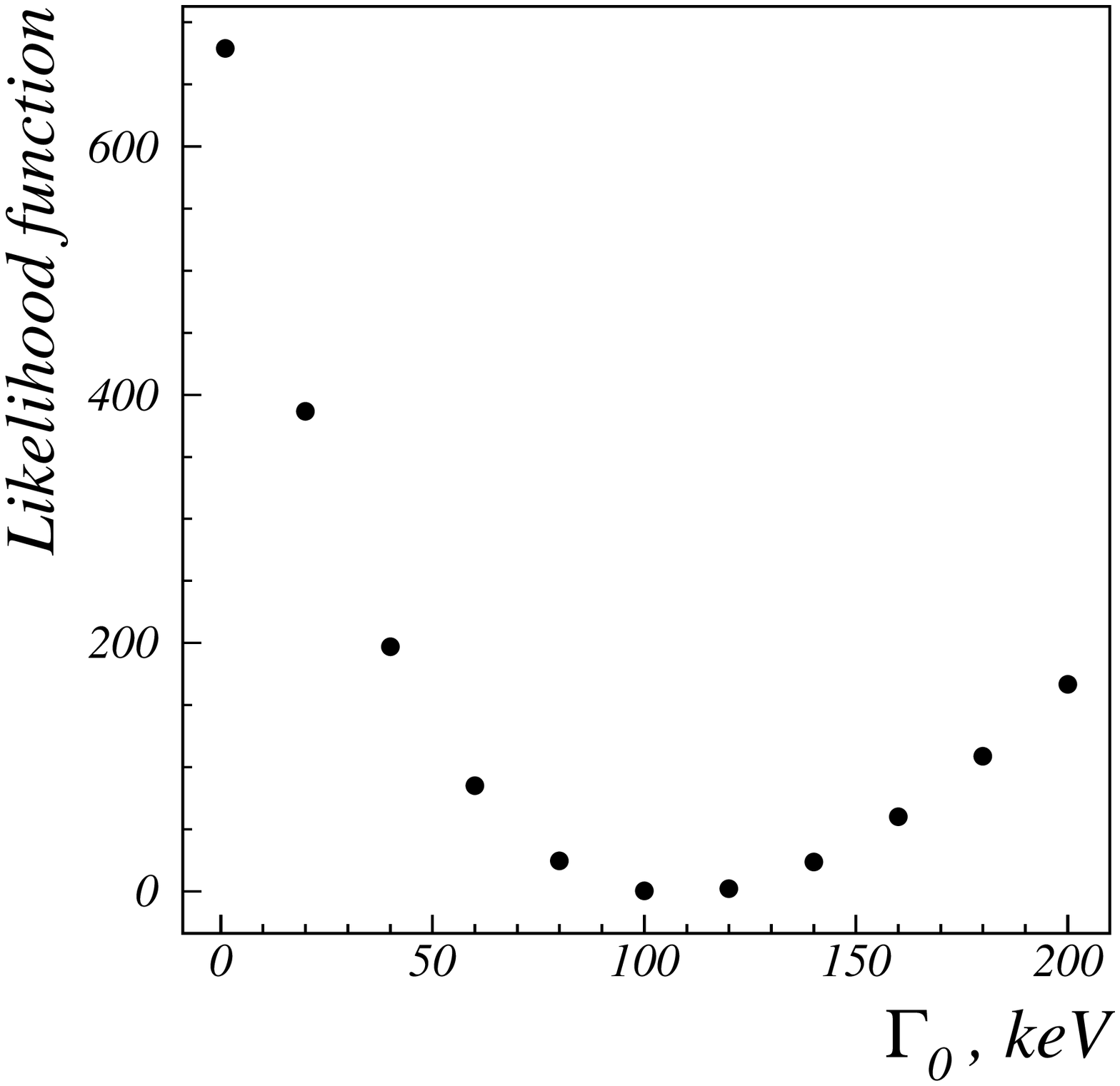} 
\end{tabular}
\caption{\label{fig:like}
Likelihood function versus measured $D^{\ast +}$ width for the nominal (left),
tracking (center), and kinematic (right) selected data samples.}
\end{figure}

Figure~\ref{fig:like}
displays the likelihood as a function of the width of the $D^{\ast +}$
for the fits to the three data samples.

The agreement is excellent among the fits to three sample, 
and when the offsets are applied we obtain
$D^{\ast +}$ widths listed in the last row in Table~\ref{tab:data}. 
The indicated uncertainties are only statistical.

%\section{Systematic Uncertainties}

	We discuss the sources of systematic uncertainties on our measurements
of the width of the $D^{\ast +}$ in
the order of their size.  The most important contribution is the variation
of the result as a function of the kinematic parameters of
the $D^{\ast +}$ decay.
	The next most important contribution
comes from any mismodeling of $\sigma_Q$'s
dependence on the kinematic parameters.  
	We take into account correlations among the
less well measured parameters of the fit, such as $k$, $f_{mis}$, and
$\sigma_{mis}$,
by fixing each parameter at $\pm 1$ standard deviation 
from their central fit values,
repeating the fit, and adding in quadrature the variation in
the width of the $D^{\ast +}$ and $Q_0$ from their central values.
	We have studied in the simulation the sources of
mismeasurement that give rise
to the resolution on the width of the $D^{\ast +}$ by replacing the
measured values with the generated values for various kinematic
parameters of the decay products.  We have then compared
these uncertainties with analytic expressions for the uncertainties.
The only source of resolution that we cannot account for in
this way is a small distortion of the kinematics of the event
caused by the algorithm used to reconstruct the $D^0$ origin point
described above. 
	We consider uncertainties from the background shape by allowing the
coefficients of the background polynomial to float. 
	Minor sources of uncertainty are from the width offsets derived
from our simulation, and our data storage digitization resolution of 1~keV.

	An extra and dominant source of uncertainty on $Q_0$ is the energy
scale of our measurements.  We are still evaluating the size of this
contribution.

	Table~\ref{tab:systematic} summarizes the systematic
\begin{table}
\caption{Systematic uncertainties on the width of the $D^{\ast +}$ and $Q_0$}
\begin{center}
\begin{tabular}{|c|c|c|c|c|c|c|} \hline
                         & \multicolumn{6}{|c|}{Uncertainties in keV} \\
                         & \multicolumn{6}{|c|}{Sample} \\ \cline{2-7}
                         & \multicolumn{2}{|c|}{Nominal}
                         & \multicolumn{2}{|c|}{Tracking}
                         & \multicolumn{2}{|c|}{Kinematic} \\ \hline

Source                    & $\delta \Gamma(D^{\ast +})$ & $\delta Q_0$
                           & $\delta \Gamma(D^{\ast +})$ & $\delta Q_0$
                           & $\delta \Gamma(D^{\ast +})$ & $\delta Q_0$ \\ \hline
Running of $Q$            & 16 & 15   & 16 & 15   & 16 & 15 \\
Mismodeling of $\sigma_Q$ & 11 & $<1$ &  9 &  4   &  7 & $<1$ \\
Fit Correlations          &  8 &  3   &  9 &  4   &  9 &  5 \\
Vertex Reconstruction     &  4 &  2   &  4 &  2   &  4 &  2 \\
Background Shape          &  4 & $<1$ &  2 & $<1$ &  2 & $<1$ \\
Offset Correction         &  2 & NA   &  6 & NA   &  3 & NA \\
%Energy Scale              & NA & ??   & NA & ??   & NA & ?? \\
Data Digitization         &  1 &  1   &  1 &  1   &  1 & 1  \\ \hline
Quadratic Sum             & 22 & 15   & 22 & 16   & 20 & 16 \\ \hline
\end{tabular}
\end{center}
\label{tab:systematic}
\end{table}
uncertainties on the width of the $D^{\ast +}$ and $Q_0$.

%\section{Conclusion}

	In summary
	we have measured the width of the $D^{\ast +}$ by studying the
distribution of the energy release in $D^{\ast +} \to D^0 \pi^+$ followed
by $D^0 \to K^- \pi^+$ decay.
With our estimate of the systematic uncertainties for each of the
three samples being essentially the same we chose to report the result
for the sample with the smallest statistical uncertainty, the minimally
selected sample, and obtain 
\begin{equation}
\Gamma(D^{\ast +}) = 96 \pm 4 \pm 22\ {\rm keV},
\label{eq:result}
\end{equation}
where the first uncertainty is statistical and the second is systematic.

This preliminary measurement is the first of the width of
the $D^{\ast +}$, and 
corresponds to a strong coupling constants \cite{Wise}
\begin{eqnarray}
g = 0.59 \pm 0.01 \pm 0.07 & {\rm or} & 
g_{D^*D\pi} = 17.9 \pm 0.3 \pm 1.9.
\end{eqnarray}
This is consistent with theoretical predictions based on HQET and
relativistic quark models, but higher than predictions based on QCD
sum rules and lattice calculations.

\section*{Acknowledgments}

We thank I.I.Bigi, A.Khodjamirian, and P.Singer for valuable discussions.
We gratefully acknowledge the effort of the CESR staff in providing us with
excellent luminosity and running conditions.
M. Selen thanks the PFF program of the NSF and the Research Corporation, 
and A.H. Mahmood thanks the Texas Advanced Research Program.
This work was supported by the National Science Foundation, the
U.S. Department of Energy, and the Natural Sciences and Engineering Research 
Council of Canada.


\section*{References}
\begin{thebibliography}{99}

\bibitem{pred} Belyaev et al. \Journal{\PRD}{51}{6177}{1995}. 
                contains a recent survey summarizing 
                and referencing previous theoretical work.
                P.~Singer, Acta Phys. Polon. B30 3849 (1999),
                J.L.~Goity and W.~Roberts JLAB-THY-00-45,
                hep-ph/0012314, and M.Di Pierro and E.Eichten,
                hep-ph/0104208 appear since that survey.
%expand this?  Ask Alan.
\bibitem{mats} J.~Bartelt {\em et al.} (CLEO Collaboration),
               \Journal{\PRL}{80}{3919}{1998}.
\bibitem{Wise} M.~Wise,  \Journal{\PRD}{45}{R2188}{1992}.
\bibitem{ACCMOR} S.~Barlag {\em et al.},
               \Journal{\PLB}{278}{480}{1992}.
\bibitem{CLEO} Y.~Kubota {\it et al.}, (CLEO Collaboration),
               \Journal{\NIMA}{320}{66}{1992}; 
               T.~Hill, \Journal{\NIMA}{418}{32}{1998}.
\bibitem{Dmix} R.~Godang {\it et al.} (CLEO Collaboration),
                         \Journal{\PRL}{84}{5038}{2000}.
\bibitem{GEANT} R.~Brun {\it et al.}, GEANT3 Users Guide, CERN DD/EE/84-1.
\bibitem{kalman} P.~Billior, \Journal{\NIMA}{225}{352}{1984}.



\end{thebibliography}
\end{document}